# Vesto Slipher and the First Galaxy Redshifts

Laird A. Thompson, Astronomy Department
University of Illinois Urbana-Champaign, lthomps@astro.illinois.edu

Reich (2011) highlights several recent discussions regarding who first discovered the velocity-distance relation among galaxies (cf., Block 2011, Nussbaumer & Bieri 2011, van den Bergh 2011a, 2011b, Way & Nussbaumer 2011), and in summary these reports show that Georges Lemaitre's work (Lemaitre 1927) was suppressed by an unknown editor of the Monthly Notices of the Royal Astronomical Society when it was translated into English as Lemaitre (1931).  Two years later Hubble (1929) found the same result.  Although the references cited above either suggest or imply that credit for this discovery be transferred from Hubble to Lemaitre, the purpose of this brief contribution is to re-emphasize the key role played by Vesto Slipher in this early stage of modern cosmology.  While Slipher never attempted to plot velocity against distance, anyone who decided to investigate this problem in the 1920's relied on Slipher's galaxy redshifts.  It was the high Doppler shifts that Slipher obtained (as high as 1800 km/s by 1921) that drew special attention to spiral galaxies and sparked further investigation into their connection to general relativity.  As described below, Arthur Eddington (1923) made a special effort to include a complete list of Slipher's redshifts in his influential book "The Mathematical Theory of Relativity" even before anyone had succeeded in deriving a velocity-distance relation.  Perhaps it is not surprising that this prominent treatment of Slipher's empirical work attracted what Block (2011) called the "Hubble eclipse".

Few historical accounts say why and how Slipher succeeded in getting 41 spiral galaxy Dopper shifts by 1921.  This note is written for the convenience of those who might be interested in how it happened but have not invested the time to figure it out.  Furthermore, few cite the original papers by Slipher (1913, 1915, 1917, 1921), and this note will make it easy for others to see the significance of his work.  Based on information provided by the SAO/NASA ADS web site, Slipher's key papers have only been cited 6, 4, 19 & 0 times, respectively.  Lemaitre (1931) has been cited 65 times and Hubble (1929) 284 times.

In a simple succinct statement, Vesto Slipher was the first to photographically detect galaxy spectra with sufficient S/N to reliably measure their Doppler shifts.  Although Milton Humason extended the galaxy redshift work to fainter galaxies on behalf of Edwin Hubble, the astronomers at Mt. Wilson would not have made rapid progress without Slipher's pioneering effort.  Hubble was fully aware of the significance and priority of Slipher's early spectroscopy (Putnam 1994), but consistent with his style of claiming sole credit for most topics he worked on, Hubble never emphasized this point.  In what follows, I will explain the early development of nebular spectrographs, briefly state Slipher's technological breakthroughs, and summarize how the Mt. Wilson group moved on to measure redshifts of even fainter galaxies.

Prior to the use of dry photographic emulsions, spectroscopy was a visual science.  Starting at about the time of Sir William Huggins in the early 1860's, spectral lines were identified in visual spectroscopes, and for nebulae, a distinction was made between emission line objects (eg. H II regions and planetary nebulae) and those that showed a faint continuous spectrum.  By the early 1900's, these were called, respectively, "green" and "white" nebulae.  Visual Doppler velocities could not be measured with high accuracy for either stars or nebulae because the wavelength shifts were small.  As astronomers learned which photographic emulsions were more light-sensitive, Doppler shifts became accessible.  Because of the way the first astronomical spectrographs were designed, stars were the first to have their Doppler shifts measured.  By the late 1800's photographic spectroscopy of stars was commonplace and "green" nebulae had been detected and studied.  It was at this point in time that Vesto Slipher began his career.

Excellent accounts of Slipher's accomplishments, as well as the events that motivated his work, are discussed in a biographical memoir of Vesto Slipher by Hoyt (1980) and in the history of Lowell Observatory as told by Putnam (1994).  Slipher arrived at Lowell Observatory in 1901 immediately after receiving his undergraduate degree in mechanics and astronomy at Indiana University and was put to work commissioning a new spectrograph.  [He returned to Indiana University in 1909 to be granted his Ph.D. degree.]  The new spectrograph had been purchased by Percival Lowell from the well-known instrument maker John Brashear, and Slipher was assigned the job of making it work on the Lowell 24-inch refractor.  As his first observing project, Percival Lowell asked Slipher to use the new spectrograph to measure the rotation velocity of Venus' atmosphere, a project that Slipher completed by 1903.  Because of frequent written correspondence between Lowell (from his Boston office) and Slipher (from the observatory in Flagstaff, AZ), a record of these developments is still available.

Believing that spiral nebulae were an early stage in the formation of other solar systems, Lowell set Slipher on a program to understand the spectra of spirals starting in 1906.  The Brashear spectrograph had been designed to use either 1, 2 or 3 prisms in series (3 prisms for the highest dispersion) in conjunction with a long focal length spectrograph camera working at f/14.2.  We know today that the speed of a spectrograph is different for point sources than it is for extended sources, and it so happened that Brashear spectrographs (also used at other major observatories in this era) could detect stellar spectra and high surface brightness "green" nebulae, but they were not of much use in the detection of galaxies because of the galaxies' lower surface brightness.

Slipher soon came to several conclusions about how to photographically detect the light from spiral nebulae.  His published papers, as well as his correspondence with Lowell, contain the following points.  First, when a nebula is placed on the slit of a spectrograph, the brightness of the sit image at the photographic plate is greatest at the lowest spectral dispersion (using 1 prism), but if all 3 prisms are used, the slit width can be widened to compensate, so the speed is not compromised (for a given velocity accuracy).  Second the speed of the spectrograph for a spiral of uniform surface brightness depends almost entirely on the f/ratio of the camera lens that focuses the slit image on the photographic plate.  The spectrograph camera must be "fast" (a small f/ratio) to shorten the exposure time.  Finally Slipher correctly concluded that for pure nebular spectroscopy, the speed of detection is independent of the telescope aperture and the telescope f/ratio. [This last statement was strictly true in the photographic era and is somewhat less so now.]  Today, these three conclusions are well known.  They were first expressed algebraically by Bowen (1952) when describing the spectrographs for the Palomar 200-inch telescope.  Bowen (1962) extends these relations.  Because galaxy nuclei are not purely "nebular", a larger telescope does work better if the spectrograph is properly designed.

By 1910 Slipher was still experimenting with and improving the Brashear spectrograph aiming to detect spiral nebulae, and he reported to Lowell that the spectrograph was working 100 times faster than before for nebular sources. By late 1912 the factor of improvement was 200 times.  The biggest gain came by replacing the original f/14.2 spectrograph camera lens with an f/2.5 lens (a commercial lens from Voigtlander).  Given that a sufficiently high S/N for a galaxy spectrum often required several nights of integration on a photographic plate (even after Slipher's improvements to the spectrograph), these modifications were absolutely crucial for Slipher's success.  Simultaneously, Slipher experimented with chemical dyes and plate baths for photographic emulsions to enhance their speed.  He also spent time observing stellar velocity standard stars so he could understand systematic errors in the spectrograph.

After making these improvements, in September 1912, Slipher used the Brashear spectrograph on the Lowell 24-inch refractor to take a 6 hour exposure of M31.  Not fully satisfied with the results, in mid-November and early December, 1912, he took two more spectra of M31, integrating each over a 2 night period.  Finally on 28 December 1912 he began a 3 night exposure of M31.  After measuring these four plates with a hand-cranked microscope-micrometer, Slipher concluded that M31 is moving towards the Sun at about 300 km/s, higher than any object known at that time.  Upon learning about this result, Percival Lowell encouraged Slipher to measure other spirals.  By the time of the 1914 meeting of the American Astronomical Society, Slipher was able to report Doppler shifts for 15 spirals.  At that point, the highest velocity nebula in his sample showed a positive Doppler shift of 1100 km/s.  In a paper published in 1917, Slipher reported Doppler shifts for 25 spirals.  In Slipher (1921) two high redshift spectra were reported, one with a Doppler velocity of 1300 km/s and the other 1800 km/s.

Arthur Eddington (1923) showed great insight in the early 1920's when, during the preparation of his widely acclaimed book "The Mathematical Theory of Relativity", he obtained from Slipher through direct correspondence a list of 41 galaxy Doppler shifts that Eddington included in his Chapter 5.  By that date, other astronomers had confirmed a number of Slipher's measurements.  Four of 41 spirals are approaching the Sun.  Eddington (1923) discusses the distribution of velocities and tries his best to make some sense of them in the context of general relativity, but he was unable to do what Lemaitre would do just four years later (Lemaitre 1927).

Slipher's spectroscopic methods were widely known, and in the 1920's the astronomers at Mt. Wilson Observatory followed Slipher's lead and began using fast spectrograph cameras in their galaxy redshift programs at the Mt. Wilson 60-inch and 100-inch telescopes.  Humason (1931) describes starting his work with a 3-inch focal length spectrograph camera that was considerably faster (f/1.43) than Slipher's f/2.5 Voigtlander camera lens.  However, images produced by the 3-inch camera were not of high quality (as reported in HMS = Humason, Mayall, and Sandage 1956).  In the late 1920's the Observatory Council at the California Institute of Technology contracted with W.B. Rayton (Rayton 1930) to make a better and even faster camera lens for the Mt. Wilson spectrograph.  The resulting f/0.59 camera lens allowed Humason to measure redshifts of significantly fainter galaxies.  This fast camera was the key to the success of the Mt. Wilson Observatory galaxy redshift program.  Hubble played no direct role in either the observational details or the development of the equipment.  HMS states that "these data are now available … chiefly as the result of Hubble's inspiring influence on his colleagues".

To remain vital, the cosmology community must recognize and reward with appropriate citations the pioneers who push the field forward. Vesto Slipher is in this group, and his contributions need to be recognized as such. Edwin Hubble was another pioneer of 20[th] century cosmology, but over and over, he made sure his legacy would be remembered. Among the methods he used was selective referencing: failing to bring attention in his publications to his competitors' best work. It is well known that Harlow Shapley had a life-long complaint against Edwin Hubble in this regard. Selective referencing has been used by other cosmologists to advance their careers in the 20[th] century. The only defense against it is to expose it, and Block (2011) has done exactly that for Hubble. The cosmology community can debate whether references to Lemaitre (1927) should now take precedence over those to Hubble (1929) for the velocity-distance relation, but it seems fair to prominently acknowledge Slipher's contributions as well.